 \documentclass[final,1p,times]{elsarticle}

\usepackage{graphicx}

\usepackage{amssymb}
\usepackage{amsthm}
\usepackage{amsmath}

\biboptions{sort&compress}

\journal{Journal of Computational Physics}

\newcommand{\be}{\begin{equation}}
\newcommand{\ee}{\end{equation}}
\newcommand{\bs}{\begin{eqnarray}}
\newcommand{\es}{\end{eqnarray}}
\usepackage{bm}

\begin{document}

\begin{frontmatter}



  \title{Unified framework for a side-by-side comparison of different
    multicomponent algorithms: lattice Boltzmann vs. phase field
    model}


\author{Luca Scarbolo\corref{cor1}\fnref{label1}}
\ead{luca.scarbolo@uniud.it} \address[label1]{Dipartimento di Energetica e Macchine, 
Universit\`a degli Studi di Udine, 33100, Udine, Italy}
\author[label2,label1]{Dafne Molin} \ead{dafne.molin@ing.unibs.it}
\address[label2]{Dipartimento di Ingegneria Meccanica, Universit\`a degli 
Studi di Brescia, 25133, Brescia, Italy} \author[label4]{Prasad Perlekar}
\ead{p.perlekar@tue.nl} \author[label3]{Mauro Sbragaglia}
\ead{sbragaglia@roma2.infn.it} \address[label3]{Dipartimento di Fisica and 
INFN, Universit\`a di Roma ``Tor Vergata'', 00133, Rome, Italy}
\author[label1]{Alfredo Soldati} \ead{soldati@uniud.it}
\author[label4]{Federico Toschi} \ead{f.toschi@tue.nl}
\address[label4]{Department of Applied Physics and Mathematics and
  Computer Science,Technische Universiteit Eindhoven, POBox 513, 5600 MB Eindhoven, Netherlands and
CNR-IAC, Via dei Taurini 19, 00185 Rome, Italy}

\cortext[cor1]{Corresponding author.}

\begin{abstract}
  Lattice Boltzmann Models (LBM) and Phase Field Models (PFM) are two
  of the most widespread approaches for the numerical study of
  multicomponent fluid systems. Both methods have been successfully
  employed by several authors but, despite their popularity, still
  remains unclear how to properly compare them and how they perform on
  the same problem. Here we present a unified framework for the
  direct (one-to-one) comparison of the multicomponent LBM against the
  PFM. We provide analytical guidelines on how to compare the
  Shan-Chen (SC) lattice Boltzmann model for non-ideal multicomponent
  fluids with a corresponding free energy (FE) lattice Boltzmann
  model. Then, in order to properly compare the LBM vs. the PFM, we
  propose a new formulation for the free energy of the
  Cahn-Hilliard/Navier-Stokes equations. Finally, the LBM model is
  numerically compared with the corresponding phase field model solved
  by means of a pseudo-spectral algorithm. This work constitute a
  first attempt to set the basis for a quantitative comparison
  between different algorithms for multicomponent fluids. We limit our
  scope to the few of the most common variants of the two most
  widespread methodologies, namely the lattice Boltzmann model (SC and
  FE variants) and the Phase Field Model.
\end{abstract}


\end{frontmatter}


\section{Introduction}
\label{Introduction}

Since the beginning of the last century several approaches have been
developed for theoretical analysis and numerical simulation of
multicomponent and multiphase flows. The most common theoretical
approach to study such complex systems is based on the sharp-interface
assumption, in which the interface between the different fluids is
considered of zero-thickness. Each component is characterized by its
own physical properties, such as density, concentration, viscosity,
etc. which are uniform over the portion of the domain occupied by the
single fluid component. These physical properties may change in a
discontinuous way across the interface and their jumps are determined
by the equilibrium conditions at the interface. The evolution of the
system is described by a set of conservation laws (mass, momentum,
energy, etc.) separately written for each component. Their solution
requires also a set of boundary conditions at the interface, which
ultimately has to be tracked. The specific numerical difficulties involved 
with the interface tracking can be circumvented by the adoption of diffuse 
interface methods, like the ones focus of the present study: the multicomponent Lattice
Boltzmann method (LBM) and the Phase Field Model (PFM). The drawback of these models is 
the high numerical resolution necessary to model real interface thickness 
that require a fictitious enlargement of the interface thickness. 
Despite this limitation, in recent years several researchers worked on the development and refinement of PFM as 
well as different variants of the multicomponent LBM.

\subsection{Introduction to LBM}

The kinetic theory for multicomponent fluids and gas mixtures has
received a lot of attention in literature
\cite{GrossJackson59,Sirovich62,Hamel65,Hamel66,Sirovich66,ZieringSheinblatt66,GoldmanSirovich67}. 
Many of the kinetic models developed for the study of mixtures are based on the 
linearized Boltzmann equations, especially the single-relaxation-time model 
due to Bhatnagar, Gross, and Krook \cite{BGK54}, also named BGK-model: 
\bs\label{EQMASTER}
\frac{\partial f({\bm x},{\bm v},t)}{\partial t}+{\bm v}\cdot {\bm \nabla} 
f({\bm x},{\bm v},t)+{\bm a} \cdot{\bm \nabla}_{\bm v} f({\bm x},{\bm v},t) = \Omega({\bm x},t)=\\
=-\frac{1}{\tau}\left[f({\bm x},{\bm v},t)-f^{(eq)}({\bm x},{\bm v},t)
\right], \nonumber \es where $f({\bm x},{\bm v},t)$ is the probability
density function to find at the space-time location $({\bm x},t)$ a
particle with velocity ${\bm v}$. The collisional kernel, on the right
hand side of Equation (\ref{EQMASTER}), stands for the relaxation
(with a characteristic relaxation time $\tau$) towards the local
equilibrium $f^{(eq)}({\bm x},{\bm v},t)$ which, in turn, depends on
the local coarse grained variables, as density and momentum: \be
\rho({\bm x},t)=\int f({\bm x},{\bm v},t) \mbox{d} {\bm
  v}\hspace{.2in} \rho {\bm u}({\bm x},t)=\int f({\bm x},{\bm v},t)
{\bm v} \mbox{d} {\bm v}.  \ee ${\bm a} \cdot{\bm \nabla}_{\bm v}
f({\bm x},{\bm v},t)$ represents the effect of a volume/body force
density, ${\bm a}$, on the kinetic dynamics. Modern discrete-velocity
counterparts of (\ref{EQMASTER}), the so-called Lattice Boltzmann
methods (LBM), are able to simulate multiphase and multicomponent
fluids and have attracted considerable attention from the scientific
community
\cite{LuoGirimaji02,LuoGirimaji03,SC_93,SC_94,SD_95,SD_96,YEO1,YEO2,YEO3,Shan11}. The
LBM is an discrete form of Boltzmann kinetic equation describing the
dynamics of a fictitious ensemble of particles
\cite{Benzi92,Gladrow00,Succi05,Chen98}, whose motion and interactions
are confined to a regular space-time lattice. This approach consists
in the following evolution: \be f_{i}({\bm x} + {\bm c}_i \Delta t, t
+ \Delta t)-f_{i}({\bm x} , t ) = -\frac{\Delta t}{\tau}[f_{i}({\bm x}
, t )-f_{i}^{(eq)}({\bm x} , t )], \ee where $f_{i}({\bm x},t)$ is the
probability density function of finding a particle at site ${\bm x}$
and time $t$, moving in the direction of the $i$-th lattice speed
${\bm c}_i$ with $i=0,\dots,b$. Systematic ways to derive the discrete
set of velocities in these models are either the discretization of the
Boltzmann equation on the roots of Hermite polynomials
\cite{Shan06,Philippi06,Siebert07,ShanHe98,Martys98,Meng11,Nie08} or
the construction of high-order lattices for more stable LBM based on
entropic approaches \cite{Chikatamarla09,Chikatamarla06}. At the same
time, the translation of the body/volume force ${\bm a} \cdot{\bm
  \nabla}_{\bm v} f({\bm x},{\bm v},t)$ onto the discrete-lattice
framework represented one of the most challenging issues in the last
years of Lattice Boltzmann research
\cite{SC_93,SC_94,SD_95,SD_96,YEO1,YEO2,YEO3,HeDoolen,MartysShanChen,HeShanDoolen,Yuan,Kupershtokh}. Through
one of the first approaches proposed in the literature, the so called
Shan-Chen (SC) approach \cite{SC_93,SC_94}, the non-ideal interactions
have been introduced directly at the discrete lattice level among the
constituent (kinetic) particles \cite{SC_93,SC_94,Shan08}. These
lattice forces embed the essential features and are able to produce
phase separation (i.e. a non-ideal equation of state and a non-zero
surface tension) as well as a detailed diffuse interface
structure. The application of the SC models has been particularly
fruitful for many applications
\cite{Hyv,Kupershtokh,ZhangTian08,Sbragaglia06,Sbragaglia09,CHEM09}.
Nevertheless, its theoretical foundations have been object of debate
in the recent years \cite{HeDoolen,Shan08,Sbragaglia06,Sbragaglia09},
mainly because of the thermodynamic consistency of the mesoscopic
interactions involved. On the other hand, in the so called free-energy
(FE) models \cite{YEO1,YEO2,YEO3}, the collisional properties of the
model have been chosen in such a way that the large scale equilibrium
is consistent with an underlying free energy functional, embedding
both hard core effects and weak interacting tails. In this case, more
traceable theoretical foundations have been provided, at least from
the point of view of a continuum theory \cite{Wagner06}. Among others,
some studies have also performed where more elaborated lattice models,
including the effect of an exclusion volume based on Enskog theory
\cite{HeDoolen,HeShanDoolen,MartysShanChen}, effective equilibria
\cite{ZhangTian08}, or even effective SC forces, were designed to
match the desired bulk pressure of a given fluid \cite{Yuan,Kupershtokh}.

\subsection{Introduction to PFM}
The phase field model is based on the idea that the interface between
two fluids is a layer of finite thickness rather than a sharp
discontinuity.  Across the interfacial layer the physical properties
of the mixture components vary in a smooth and continuous way, from
one fluid to the other. This approach is based on the pioneering work
of van der Waals \cite{WDW}, who first determined the interface
thickness of a critical liquid-vapor mixture.  In the PFM the state of
the system is described, at any time, by an order parameter
$\phi=\phi(\bm x)$, which is a function of the position vector ${\bm
  x}$.  The order parameter directly represents a physical properties
of the fluid, such as its density, molar concentration, etc.; all the
remaining properties are in turn modeled as proportional to $\phi(\bm
x)$ \cite{ANDERSON05,BCB}. According to the diffuse-interface
approach, the order parameter is continuous over the entire domain and
it shows smooth variations across the interface between single fluid
regions, where $\phi(\bm x)$ assumes approximately uniform
values. Coupling the continuous and diffuse representation of the
system with a transport equation of the order parameter, the system
evolution can be resolved in time.  One of the best-known PFM is the
Cahn-Hilliard equation \cite{CHANHILL,CHANHILL2}, an extension to
binary mixtures of the work of Van der Waals \cite{WDW}. This equation
is a transport equation for the order parameter, where the evolution
of $\phi(\bm x)$ is proportional to the gradient of the chemical
potential, $\mu$. The chemical potential is defined in terms of the
free energy functional, ${\cal F}[\phi]$: \be\label{eq:chemgeneraldef}
\mu=\frac{\delta{\cal F}[\phi]}{\delta \phi}, \ee where the free
energy, ${\cal F}[\phi]$, assumes suitable definitions according to
the problem under analysis (and also depending on which physical
quantity has to be described) and is a conservative, thermodynamically
consistent functional. The most common free energy formulation is
given by the sum of an ideal part, ${\cal F}_{id}[\phi]$, and a
non-local part, ${\cal F}_{nl}[{\bm \nabla} \phi]$. The ideal part
accounts for the tendency of the system to separate in two pure
components and is derived from the thermodynamics of mixtures. The
non-local part accounts for the diffusive interfacial region. Through
the Cahn-Hilliard equation, the evolution of the order parameter is
thermodynamically consistent and subject to a phase field
conservation. As a result, the prediction of the interfacial layer
does not deteriorate.  In the case of density-matched fluid systems,
the convective Cahn-Hilliard equation is coupled with a modified
Navier-Stokes equation, where a surface tension (or capillary) forcing
term, which is derived from the Korteweg stress \cite{KORTEWEG}, is
introduced.  This contribution yields to the
Cahn-Hilliard/Navier-Stokes coupled equations system
\cite{JAQMIN99,LW_TUR,KAWASAKI70}, which is also known as Model-H,
according to the classification of Hohenberg and Halperin
\cite{HOEHAL}, who studied the convective phase separation of a
partially miscible fluid mixture.  This model, originally developed to
study critical phenomena
\cite{Mauri_Shinnar,Mauri_Vladimirova,Molin_Mauri,Molin2,Lamorgese},
was subsequently used by many authors to study two-phase flows of
Newtonian fluids \cite{ANDERSON98,JAQMIN99,LIUSHEN03,BCB,YUE}. In this
case, even if the fluids are in fact immiscible, molecular diffusion
between the two species is allowed in the interfacial region. Thus,
the thinner is the interface, the more realistic is the numerical
solution. Realistic interface thickness requires high numerical
resolution, which is usually beyond current computational limits. For
this reason the interface is often kept larger than corresponding
physical value (this approximation holds also for lattice Boltzmann 
approaches). However, in spite of this approximation, the method has shown 
capabilities to capture complex interfacial dynamics in a wide range of real 
physical problems \cite{ANDERSON05,JAQMIN99,JAQMIN00,YUE10}.

\subsection{Advantages and disadvantages of the methods}
Multicomponent LBM and PFM have demonstrated excellent performances to
predict the dynamics of multiphase and multicomponent flows. Yet, both
methods show their own peculiar characteristics and drawbacks which
can limit their use, performances and range of validity.  A particular
unexpected, and unwanted, feature of multiphase and multicomponent
solvers is the manifestation of non physical velocities near
equilibrium interface, present even for systems at rest. From a
physical viewpoint the velocity should clearly vanish at equilibrium
but, as it has been observed by many authors, small spurious currents
most often exist in the proximity of the interfaces. In an attempt to
remove these unwanted features several improvement to the LBM have
been proposed in recent years
\cite{GUO11,CASTEA03,WAGNER03,LEE06,SHAN06,SBRAG07}. It worths
noticing that some of these improvement are capable to remove these
spurious current to machine precision \cite{LEE06}.  Spurious currents
have also been observed in other numerical methods including the PFM
\cite{JAME02,RENARDY02,SCARDOVELLI99}.

Because of the magnitude of these spurious current drastically depend
on the actual variant of the LBM or PFM, the comparison between the
two methods may be somewhat ill defined. Here we aim at comparing the
LBM vs. the PFM for their most widespread and used variants. Our
answer will thus not provide a general statement valid for the two
methods as such, but will still provide some extremely useful insight
in what can be expected from the most employed variations of the
approaches.  As a side results, we will also quantitatively compare 
two of the most widely used lattice Boltzmann variants, the SC-LBM and
the FE-LBM, under the same conditions (i.e. same diffuse interface
model, same surface tension, same chemical potential, etc.). In order
to achieve our goal, the problem of a one-to-one matching of the PFM
with SC/FE multicomponent LBM needs to be addressed first.
The one-to-one matching of the two methods gives also the opportunity to
clarify how they compare with respect to the computational costs.
In order to address these issues we start by analyzing the SC model
for two population with inter-particle repulsion; the large scale
continuum limit is reviewed and formulated in terms of a diffuse
interface model with an underlying thermodynamic FE functional. In
this way one of the crucial issues in the matching of SC model vs.
the corresponding FE model is being solved. Then, starting from the
matched SC/FE multicomponent LBM, a new formulation for the
free-energy of the PFM is derived in order to directly compare
them. Finally, a comparison of the numerical results obtained, on the
same problem, from both LBM and PFM is presented, focusing in
particular on unwanted spurious currents or mass leakage in sheared
suspensions. This work focuses only the case of binary mixtures, 
even if modeling more than two components is nowadays far from trial 
extension \cite{Arcidiacononew,Asinarinew,Shannew,KIM_LOW,PARK_MAUR_AND}.

\section{The lattice Boltzmann models}
\label{The lattice Boltzmann methods}

\subsection{The multicomponent Shan-Chen model}
\label{The multicomponent Shan-Chen (SC) model}

In this section the multicomponent model introduced by Shan \& Chen
\cite{SC_93,SC_94} is reviewed. First the main properties of the model
are recalled, then the equilibrium features (diffusive current and
pressure tensor) relevant on the hydrodynamic scales are
analyzed. Starting from a kinetic lattice Boltzmann equation
\cite{Benzi92,Gladrow00,Chen98} for a multicomponent fluid with $N_s$
species \cite{SD_95,SD_96}, the evolution equations over a
characteristic time lapse $\Delta t$ read as follows: \be\label{eq:be} f_{is}({\bm
  x} + {\bm c}_i \Delta t, t + \Delta t)-f_{is}({\bm x} , t ) =
-\frac{\Delta t}{\tau_s}[f_{is}({\bm x} , t )-f_{is}^{(eq)}(\rho_s ,
{\bm u}+\tau_s {\bm F}_{s} /\rho_s)], \ee where
$f_{is}({\bm x},t)$ is the probability density function of finding a
particle of species $s=1,\dots,N_s$. For the sake of simplicity, the
characteristic time interval $\Delta t$ is assumed to be unity in what
follows. The left hand-side of (\ref{eq:be}) accounts for the
molecular free-streaming, whereas the right-hand side represents the
time relaxation (due to collisions) towards local Maxwellian
equilibrium $f_{is}^{(eq)}(\rho_s, {\bm u})$ on a time scale $\tau_s$
\cite{Benzi92,Chen98,Gladrow00,BGK54}. The local Maxwellian is
truncated at second order, leading to a sufficiently accurate
approximation to correctly recover the hydrodynamic balance of an
isothermal regime \cite{Benzi92,Gladrow00,Succi05,Chen98}.  It should
be noted that the equilibrium for the $s$ species is a function of
their local densities and common velocities, defined as: \be
\rho_s({\bm x}, t)=\sum_i f_{is}({\bm x}, t); \hspace{.2in} {\bm u}
({\bm x}, t)= \frac{\sum_s \frac{1}{\tau_s} \sum_i f_{is}({\bm x}, t)
  {\bm c}_i }{ \sum_s \frac{1}{\tau_s} \rho_s ({\bm x}, t)}. \ee This
common velocity receives a suitable shift from the force, ${\bm
  F}_{s}$, acting on the $s$ species \cite{SC_93,SD_95}. ${\bm F}_{s}$
may be either an external force or an internal force representing an
intermolecular interactions:
\begin{eqnarray}\label{FORCE} {\bm F}_s ({\bm x},t) = -G_{s s'} \rho_s
  ({\bm x},t) \sum_{s' \neq s}\sum_{i} w_i^{(eq)} \rho_{s'}({\bm
    x}+{\bm c}_i,t) {\bm c}_i,
\end{eqnarray}
where $w_i^{(eq)}$ are suitable weights used to enforce isotropy of
the resulting hydrodynamics \cite{Benzi92,Gladrow00,Chen98}. The
weights, $w_i^{(eq)}$, are normalized as follows: \be \sum_i
w_i^{(eq)} c_i^a c_j^b=\delta_{ab} c_S^2, \ee \be \sum_i w_i^{(eq)}
c_i^a c_i^b c_i^s c_i^k=c_S^4\left(\delta_{ab} \delta_{sk}+\delta_{ak}
  \delta_{bs}+\delta_{as} \delta_{bk} \right), \ee where $\delta_{ab}$
is the Kronecker delta and $c_S^2=1/3$ a constant. In the
long-wavelength limit (i.e. when the fluctuations with respect to the
equilibrium distribution function are small) the set of macroscopic
equations associated with the kinetic model consist of the continuity
equations (one for each component) and of the equation of motion for
the total fluid momentum. Dealing with the two species (i.e. $A$ and
$B$) under the assumption of the same characteristic time scale for
all the components $\tau_s=\tau$,\footnote[1]{Whenever timescales are
  different, the characteristic time maps directly onto an effective
  relaxation time. For the two species system $(A, B)$ this assumes
  the form $\bar \tau= \frac{\rho_A \tau_A+\rho_B \tau_B}{\rho}$;
  similar expressions need to be used in the total baricentric
  velocity.}  these equations are approximated by: 
\be \frac{\partial
  \rho_s}{\partial t} + {\bm \nabla} \cdot (\rho_s {\bm u}) = {\bm
  \nabla} \cdot {\bm J}^{(s)}, 
\ee 
\be 
\rho \left( \frac{\partial {\bm
      u}}{\partial t}+{\bm u} \cdot {\bm \nabla} {\bm u}\right)=-{\bm \nabla} \cdot {\bm P}+{\bm \nabla}\cdot(\eta(\tau) {\bm \nabla} {\bm
  u}+\eta(\tau) {\bm \nabla} {\bm u}^T),
\ee 
where $\rho=\sum_{s} \rho_s$ is the total density, ${\bm u} =
\sum_s \rho_s {\bm u}_s/\rho$ is the baricentric (total) fluid
velocity and ${\bm P}$ the pressure tensor with the property $-{\bm
  \nabla} \cdot {\bm P}+{\bm \nabla} (c_S^2 \rho)=\sum_s {\bm F}$. The kinematic viscosity is $\nu(\tau)=c_S^2\left(\tau-1/2\right)$ and the dynamic viscosity is $\eta(\tau)=\rho \nu(\tau)$. The diffusive current, ${\bm J}^{(s)}$,
is given by the following: 
\bs {\bm J}^{(A)} & = & \frac{\rho_A
  \rho_B}{\rho} \left[c_S^2 \left(\tau-\frac{1}{2} \right)\left(
    \frac{{\bm \nabla} \rho_A}{\rho_A}-\frac{{\bm \nabla}
      \rho_B}{\rho_B} \right) -\tau \left(\frac{\bm
      F^{(A)}}{\rho_A}-\frac{{\bm F^{(B)}}}{\rho_B} \right) \right]\\
& = & -{\bm J}^{(B)}, \nonumber \es and, for purely repulsive fluids,
${\bm F}^{(A,B)}/\rho_{A,B}$ yields: \be \frac{{\bm
    F}^{(A,B)}}{\rho_{A,B}}=-g_{AB} c_S^2{\bm
  \nabla}\rho_{B,A}-\frac{g_{AB}}{2}c_S^2{\bm
  \nabla}\Delta\rho_{B,A}+\dots, \ee where the coupling constant $G_{s
  s'}=g_{AB}>0$. Both continuity equations may be subtracted,
obtaining: \be \frac{\partial \phi}{\partial t}+{\bm \nabla}\cdot
(\phi {\bm u})={\bm \nabla} \cdot {\bm J}^{(\phi)}, \ee where
$\phi=\rho_A-\rho_B$ and where ${\bm J}^{(\phi)}=2 {\bm J}^{(A)}$. The
pressure tensor in the (global) momentum equation is: \bs
{\bm P} & = &  c_s^2\left[(\rho_A+\rho_B) + g_{AB} \rho_A \rho_B  + c_S^2 \frac{g_{AB}}{2} \rho_A \Delta \rho_B +  c_S^2 \frac{g_{AB}}{2} \rho_B \Delta \rho_A \right] {\bm \delta} \\
& + & \left( {\bm \nabla} \rho_A \cdot {\bm \nabla} \rho_B\right) {\bm
  \delta} - c_S^4 g_{AB} {\bm \nabla} \rho_A {\bm \nabla} \rho_B+{\bm
  K}^{(\tau)}, \nonumber \es where the extra spurious $\tau$-dependent
contribution ${\bm K}^{(\tau)}$ is small for $\tau$ close to
$1/2$. Its exact expression is reported in \cite{CHEM09} and, for the
sake of simplicity, it will be neglected by setting $\tau$ close to
$1/2$, thus ${\bm K}^{(\tau)} \rightarrow 0$. Summarizing, the
following equations system is found: \bs
{\bm J}^{(\phi)} & = & \frac{2 \rho_A \rho_B c_S^2}{\rho}\left(\tau-\frac{1}{2} \right)\left( \frac{{\bm \nabla} \rho_A}{\rho_A}-\frac{{\bm \nabla} \rho_B}{\rho_B} \right) \\
& - & \frac{2 \rho_A \rho_B \tau}{\rho} \left[g_{AB} c_S^2 {\bm
    \nabla}(\rho_B-\rho_A)+\frac{g_{AB}}{2}c_S^2{\bm \nabla}
  \Delta(\rho_B-\rho_A) \right],\nonumber \es \bs {\bm P} &
= & c_s^2 \left[(\rho_A+\rho_B) + g_{AB} \rho_A \rho_B + c_S^2
  \frac{g_{AB}}{2} \rho_A \Delta \rho_B + c_S^2 \frac{g_{AB}}{2}
  \rho_B \Delta \rho_A \right] {\bm \delta} \\ \nonumber & + &
\left({\bm \nabla} \rho_A \cdot {\bm \nabla} \rho_B \right) {\bm
  \delta} -c_S^4 g_{AB} {\bm \nabla} \rho_A {\bm \nabla} \rho_B.  \es
The variables $\rho=\rho_A+\rho_B$ and $\phi=\rho_A-\rho_B$ can be
introduced such as $\rho_A=(\rho+\phi)/2$ and
$\rho_B=(\rho-\phi)/2$. Under the assumption of an incompressible
fluid at constant $\rho$, the derivative with respect to the density
can be neglected in the pressure tensor: \bs\label{PT}
{\bm P} & = & c_s^2 \left[\rho + \frac{g_{AB}}{4}(\rho^2-\phi^2)  - c_S^2 \frac{g_{AB}}{4}\phi \Delta \phi -c_s^2 \frac{g_{AB}}{8}|{\bm \nabla} \phi|^2 \right] {\bm \delta} \\
& + & c_s^4 \frac{g_{AB}}{4}{\bm \nabla}\phi {\bm \nabla} \phi,
\nonumber \es while the diffusive current is mapped into the following
expression: \be {\bm J}^{(\phi)}=c_S^2\frac{2\rho_A
  \rho_B}{\rho}\left(\tau-\frac{1}{2} \right) \left[ {\bm \nabla} \log
  \left(\frac{\rho+\phi}{\rho-\phi}\right) -g^{(\tau)}_{AB}{\bm
    \nabla} \phi - \frac{g^{(\tau)}_{AB}}{2} c_S^2 {\bm \nabla}\Delta
  \phi \right], \ee or, in a more compact form: \bs\label{Jphi}
{\bm J}^{(\phi)} & = & M(\phi)\left[\frac{c_S^2}{2} {\bm \nabla} \log \left(\frac{\rho+\phi}{\rho-\phi}\right)  - \frac{g^{(\tau)}_{AB}}{2} c_S^2 {\bm \nabla} \phi - \frac{g^{(\tau)}_{AB}}{4} c_S^4 {\bm \nabla}\Delta \phi  \right] \\
& = &M(\phi) {\bm \nabla} \mu(\phi). \nonumber \es Into the equation
above, the density-dependent mobility has been introduced: \be
M(\phi)=\frac{4\rho_A \rho_B}{\rho}\left(\tau-\frac{1}{2}
\right)=\frac{(\rho^2-\-\phi^2)}{ \rho}\left(\tau-\frac{1}{2} \right),
\ee and the chemical potential yields: \be\label{CHEMSC}
\mu^{(\tau)}(\phi)=\frac{c_S^2}{2} \log
\left(\frac{\rho+\phi}{\rho-\phi}\right) - \frac{g^{(\tau)}_{AB}}{2}
c_S^2 \phi - \frac{g^{(\tau)}_{AB}}{4} c_S^4 \Delta \phi, \ee where
the modified (through the relaxation time $\tau$) coupling constant
reads: \be g^{(\tau)}_{AB}=\frac{\tau}{(\tau-\frac{1}{2})}g_{AB}, \ee
which reduces to $g_{AB}$ in the limit $\tau \gg 1/2$. Summarizing,
the set of incompressible equations which are approximated on a large
scale are given by 
\be\label{EQ2} 
\frac{\partial \phi}{\partial
  t}+{\bm \nabla}\cdot (\phi {\bm u})={\bm \nabla} \cdot {\bm
  J}^{(\phi)}, \ee \be\label{EQ3} \rho\left( \frac{\partial {\bm
      u}}{\partial t}+{\bm u} \cdot {\bm \nabla} {\bm u}\right)=-{\bm \nabla} \cdot {\bm P}+{\bm \nabla}\cdot(\eta(\tau){\bm \nabla} {\bm
  u}+\eta(\tau){\bm \nabla} {\bm u}^T), 
\ee 
where
${\bm P}$ and ${\bm J}^{(\phi)}$ are given in (\ref{Jphi}) and
(\ref{PT}). It worths noticing that, for the models here considered,  
the phase diffusion is dictated by the same relaxation time ruling 
the kinematic viscosity, i.e. fixed Schmidt number is considered. 
In order to decouple these two time scales a different more elaborated 
formulation based on multiple relaxation time schemes (MRT) \cite{CHAI,YUetal} 
should be adopted.

\subsection{The multicomponent free energy model}
\label{The Multicomponent Free Energy (FE) model}
The other lattice Boltzmann variant that we consider in this work is
the Free Energy (FE) based model \cite{YEO1,YEO2,YEO3}. Within this
approach the equilibrium properties of the model are constructed to be
consistent with an underlying continuum free energy functional of the
order parameter, whose formulation is the following:
\be\label{FEfunctional} {\cal F}[\phi] = {\cal F}[\phi]_{id} + {\cal
  F}[{\bm \nabla} \phi]_{nl} = \int
\left[V(\phi)+\frac{\kappa}{2}|{\bm \nabla} \phi|^2
\right]\mbox{d}{\bm x}, \ee where $V(\phi)$ is the bulk free energy
density and $\kappa$ a constant parameter associated to the surface
tension at the interface. The chemical potential $\mu(\phi)$ and the
pressure tensor are derived from thermodynamics identities: \be
\mu(\phi)=\frac{\delta{\cal F}[\phi]}{\delta \phi}=\frac{\partial
  V(\phi)}{\partial \phi}-\kappa \Delta \phi, \ee \be {\bm P} =
\left(P_{b}(\phi) - \kappa \phi \Delta \phi -\frac{\kappa}{2}|{\bm
    \nabla} \phi|^2 \right) {\bm \delta}+\kappa {\bm \nabla}\phi {\bm
  \nabla}\phi, \ee where $P_b(\phi)=\phi \partial V(\phi)/\partial
\phi-V(\phi)$ is the bulk pressure. For the out of equilibrium
properties, the mesoscopic dynamics is given by the following two
coupled kinetic equations:
\begin{eqnarray}\label{FE1} f_{i}({\bm x} + {\bm c}_i
\Delta t, t + \Delta t)-f_{i}({\bm x} , t ) & = & -\frac{\Delta
  t}{\tau}[f_{i}({\bm x} , t )-f_{i}^{(eq)}(\rho , \phi, {\bm u})],\\
\label{FE2} g_{i}({\bm x} + {\bm c}_i \Delta t, t + \Delta
t)-g_{i}({\bm x} , t ) & = & -\frac{\Delta t}{\tau_g}[g_{i}({\bm x} , t
)-g_{i}^{(eq)}(\phi , {\bm u})],\end{eqnarray}
where the macroscopic density and velocity are given by: \be \rho({\bm
  x}, t)=\sum_i f_{i}({\bm x}, t); \hspace{.2in} {\bm u} ({\bm
  x}, t)= \frac{ \sum_i f_{i}({\bm x}, t) {\bm c}_i }{\rho ({\bm x},
  t)}.  \ee
The order parameter density, $\phi$, is obtained from the populations,
$g_i$: \be \phi({\bm x}, t)= \sum_i g_{i}({\bm x}, t).  \ee The
equilibrium distribution function is then chosen with a polynomial
form in the kinetic velocities, so that: \be\label{EQFE1} \sum_i
f_i^{(eq)}{\bm c}_i {\bm c}_i={\bm P}+\rho {\bm u}{\bm u}, \ee
\be\label{EQFE2} \sum_i g_i^{(eq)}{\bm c}_i {\bm c}_i={\bm \delta}
\Gamma \mu +\phi {\bm u}{\bm u}, \ee where $\Gamma$ controls the order
parameter mobility $M=\Gamma \left(\tau_g-1/2\right)$ \footnote{in the free 
energy approach the mobility can be easily changed by properly tuning both 
$\tau_g$ and $\Gamma$.}. 
The previous kinetic model approximates the following macroscopic equations 
in the incompressible limit: 
\be\label{CONTYEO} 
\frac{\partial
  \phi}{\partial t}+{\bm \nabla}\cdot (\phi {\bm u})=M \Delta
\mu(\phi), \ee \be\label{MOMYEO} \rho\left( \frac{\partial {\bm
      u}}{\partial t}+{\bm u} \cdot {\bm \nabla} {\bm u}\right)=-{\bm \nabla} \cdot {\bm P}+{\bm \nabla}\cdot(\eta(\tau){\bm \nabla} {\bm
  u}+\eta(\tau){\bm \nabla} {\bm u}^T), 
\ee 
where the viscosity is the same as in (\ref{EQ3}). The presence of extra
spurious terms in the large scale expansions
(\ref{CONTYEO})-(\ref{MOMYEO}) has been discussed in
\cite{IGNACIO}. Through the numerical simulations presented in this
work, these terms are directly evaluated and found to be negligibly
small.

\subsection{Matching in the long-wavelength limit}
\label{Matching in the Long-Wavelength limit}
In this section the details of the matching between equations
(\ref{EQ2})-(\ref{EQ3}) and (\ref{CONTYEO})-(\ref{MOMYEO}) are
reported, representing the large scale limits of the SC and FE models,
respectively, when compressible effects are negligible ($\rho$
constant). Since the transport properties in the hydrodynamical
equations are in fact the same, the crucial issue for the matching is
to set the same thermodynamical background functional.  First the
continuity equation (\ref{EQ2}) is rewritten in a formulation similar
to equation (\ref{CONTYEO}): \be\label{EQcontmatching} \frac{\partial
  \phi}{\partial t}+{\bm \nabla}\cdot (\phi {\bm u})={\bm \nabla}
\cdot {\bm J}^{(\phi)}={\bm \nabla} \left[ M(\phi){\bm \nabla}
  \mu\right], \ee where the chemical potential can be derived from the
following $\tau-$dependent functional: \bs\label{FEtau}
{\cal F}[\phi]^{(\tau)} & = & c_S^2\left(\frac{\rho+\phi}{2}\right)\log \left(\frac{\rho+\phi}{2} \right)+c_S^2\left(\frac{\rho-\phi}{2}\right)\log \left(\frac{\rho-\phi}{2} \right)  \\
& + & g^{(\tau)}_{AB}c_S^2
\frac{(\rho^2-\phi^2)}{4}+\frac{g^{(\tau)}_{AB}c_S^4}{8}|{\nabla}\phi|^2,
\nonumber \es from which, by definition, the chemical potential is
extracted: \bs
\mu^{(\tau)}(\phi) & = &\frac{\delta {\cal F}^{(\tau)}[\phi]}{\delta \phi}=\frac{\partial {\cal F}^{(\tau)}[\phi]}{\partial \phi}-{\bm \nabla} \cdot \left[\frac{\partial {\cal F}[\phi]}{\partial ({\bm \nabla}\phi)} \right] \\
& = & \frac{c_S^2}{2}\log \left(\frac{\rho+\phi}{\rho-\phi}\right)
-\frac{g^{(\tau)}_{AB}}{2} c_S^2 \phi - \frac{g^{(\tau)}_{AB}}{4}
c_S^4 \Delta \phi, \nonumber \es and it coincides with equation
(\ref{CHEMSC}). In order to set an accurate matching between the two
lattice Boltzmann approaches, two more issues must be
analyzed. Although one may think to integrate the FE model
(\ref{FE1})-(\ref{FE2}) with the functional (\ref{FEtau}) instead of
(\ref{FEfunctional}), both continuity equations are not exactly the
same due to the $\phi$ dependency of the mobility in equation
(\ref{EQcontmatching}). However, equation (\ref{EQcontmatching}) can
be approximated with the mobility in the center of the
interface (that means negligibly small density changes $\phi \ll \rho$):
\footnote{Equation (\ref{EQcontmatching}) has been simulated
  with and without the constant mobility by adding proper
  counter-terms to the lattice Boltzmann equation. No substantial
  differences have been observed for the features analyzed in this
  paper.}  
\be M(\phi,\rho,\tau) \approx
M(\rho,\tau)=\rho\left(\tau-\frac{1}{2} \right).  
\ee 
With this approximation, the advection-diffusion equation for the order
parameter $\phi$ obtained from the SC model and from equations
(\ref{FE2}) and (\ref{EQFE2}) of the FE model with the functional
(\ref{FEtau}) are set to be the same in the hydrodynamical limit.
Furthermore the pressure tensor should be matched between the two (FE
and SC) LBM descriptions. The functional of equation (\ref{FEtau})
contains a spurious and nonphysical $\tau$ dependence hidden inside
the definition of $g^{(\tau)}_{AB}=(\tau g_{AB})/(\tau-1/2)$. Only
towards the limit of large $\tau$ the functional (\ref{FEtau}) becomes
independent of $\tau$, whereas, for finite $\tau$, this extra spurious
dependency should be kept inside equation (\ref{FEtau}) in order to
accurately match both SC and FE descriptions. This difference 
emerges in the SC model because the collisional kernel does not 
conserve momentum locally \cite{SD_95,SD_96}. Therefore the lattice 
momentum should be shifted by a factor ${\bm F}\Delta t/2$ to 
reproduce the hydrodynamical equations. It is important to remark here 
that such a spurious contribution may be removed by curing the discrete 
forcing effects as described in \cite{GUO02}. This technique has already 
been adopted in recent simulations with the multicomponent SC model in 
the framework of the so-called multiple relaxation time schemes 
(MRT) \cite{CHAI,YUetal}\footnote{Since the goal of the present work 
is to propose guidelines to match the SC and the FE approaches 
we limited the analysis to the basic versions of the algorithms.}. 
Differently with respect to the chemical potential, the pressure tensor of the SC description can be directly extracted: 
\bs\label{FEcont}
{\cal F}[\phi] & = &  c_S^2\left(\frac{\rho+\phi}{2}\right)\log \left(\frac{\rho+\phi}{2} \right)  +  c_S^2 \left(\frac{\rho-\phi}{2}\right)\log \left(\frac{\rho-\phi}{2} \right)\\ & + & g_{AB}c_S^2 \frac{(\rho^2-\phi^2)}{4}+\frac{g_{AB}c_S^4}{8}|{\bm\nabla}\phi|^2 \nonumber \\
& = & {\cal F}[\phi, \rho, {\bm \nabla}\phi], \nonumber \es that is
the large $\tau$ limit of (\ref{FEtau}). In fact, following standard
procedures, the conserved current (pressure tensor) under the
hypothesis of translational invariance of the free energy can be
derived. The general expression is: \be {\bm P}=-{\cal
  F}^{\lambda_{1},\lambda_2}{\bm \delta}+\frac{g_{AB}c_S^4}{4}{\bm
  \nabla} \phi {\bm \nabla} \phi, \ee where the mass conserving free
energy has been used: \be {\cal
  F}^{\lambda_{1},\lambda_2}[\phi,\rho,{\bm\nabla}\phi]={\cal
  F}[\phi,\rho,{\bm \nabla}\phi]-\lambda_1 \phi-\lambda_2 \rho, \ee
$\lambda_{1,2}$ are two Lagrange multipliers introduced to ensure the
global conservation of $\phi$ and $\rho$. Their values are readily
evaluated using the Euler-Lagrange equations (should be noticed that
there are no gradients of $\rho$ in the free energy functional): \be
\frac{\partial {\cal F}^{\lambda_{1},\lambda_2}}{\partial \phi}={\bm
  \nabla} \cdot \left[ \frac{\partial {\cal
      F}^{\lambda_{1},\lambda_2}}{\partial ({\bm \nabla}\phi)}
\right], \hspace{.4in} \frac{\partial{\cal F}
  ^{\lambda_{1},\lambda_2}}{\partial \rho}={\bm \nabla} \cdot \left[
  \frac{\partial {\cal F}^{\lambda_{1},\lambda_2}}{\partial ({\bm
      \nabla} \rho)} \right]=0, \ee this yields: \be
\lambda_1=\frac{\partial V(\phi,\rho)}{\partial
  \phi}-\frac{g_{AB}c_S^4}{4} \Delta \phi, \hspace{.4in}
\lambda_2=\frac{\partial V(\phi,\rho)}{\partial \rho}.  \ee At the
end, the following pressure tensor is obtained: \bs
{\bm P} & = & \left[\phi \frac{\partial V(\phi,\rho)}{\partial \phi}+\rho \frac{\partial V(\phi,\rho)}{\partial \rho}-V(\phi,\rho) - \kappa \phi \Delta \phi -\frac{\kappa}{2}|{\bm \nabla} \phi|^2 \right] {\bm \delta} \\
& + & \kappa {\bm \nabla} \phi {\bm \nabla} \phi, \nonumber \es where
$\kappa=(c_S^4 g_{AB})/4)$ and the bulk potential $V(\phi)$ given by:
\bs
V(\phi,\rho) & = & c_S^2 \left(\frac{\rho+\phi}{2}\right)\log \left(\frac{\rho+\phi}{2} \right)+c_S^2 \left(\frac{\rho-\phi}{2}\right)\log \left(\frac{\rho-\phi}{2} \right) \\
& + & g_{AB}c_S^2 \frac{(\rho^2-\phi^2)}{4}. \nonumber \es It can be
immediately noticed: \be \phi \frac{\partial V(\phi,\rho)}{\partial
  \phi}+\rho \frac{\partial V(\phi,\rho)}{\partial
  \rho}-V(\phi,\rho)=c_s^2 \rho+c_S^2 \frac{g_{AB}}{4}(\rho^2-\phi^2),
\ee and, at the end: \be {\bm P}=\left(c_s^2 \rho+c_S^2
  \frac{g_{AB}}{4}(\rho^2-\phi^2) - \kappa \phi \Delta \phi
  -\frac{\kappa}{2}|{\bm \nabla} \phi|^2 \right) {\bm \delta}+\kappa
{\bm \nabla}\phi {\bm \nabla} \phi, \ee that coincides with
(\ref{PT}). We conclude that the momentum equation from the SC model
and from equations (\ref{FE1}) and (\ref{EQFE1}) of the FE model with
the functional (\ref{FEcont}) are set to be the same in the
hydrodynamical limit.

\section{The phase field model}
\label{The phase field model}

In this section the phase field model for a multicomponent fluid
system is analyzed. Starting from the features of the PFM proposed by
Badalassi {\em et al.} \cite{BCB} and Yue {\em et al.} \cite{YUE}, a
new formulation of the free energy functional is derived to match the
SC/FE multicomponent LBM reported in Section \ref{The lattice
  Boltzmann methods}.  For an isothermal binary mixture, where $\phi$
is the relative concentration of the mixture components, the free
energy functional reads: \be {\cal F}[\phi] = {\cal F}[\phi]_{id} +
{\cal F}[{\bm \nabla} \phi]_{nl} = \int
\left[V(\phi)+\frac{\kappa}{2}|{\bm \nabla} \phi|^2
\right]\mbox{d}{\bm x}, \ee where $V(\phi)$ is the ideal part of the
specific free energy, $(\kappa/2) |{\bm \nabla} \phi|^2$ is the
non-local part of the specific free energy and $\kappa$ is a constant
positive parameter associated to the surface tension at the
interface. Recalling equation (\ref{eq:chemgeneraldef}) the chemical
potential $\mu(\phi)$ is: \be\label{eq:chempdef}
\mu(\phi)=\frac{\delta {\cal F}[\phi]}{\delta \phi}= \frac{\partial
  {\cal F}[\phi]}{\partial \phi} - {\bm \nabla}\cdot
\left[\frac{\partial {\cal F}[\phi]}{\partial ({\bm
      \nabla}\phi)}\right] = \frac{\partial V(\phi)}{\partial
  \phi}-\kappa \Delta \phi.  \ee For unsteady problems, the evolution
of the system is described by the Cahn-Hilliard equation
\cite{CHANHILL,CHANHILL2}, in which the evolution of $\phi({\bm x})$
in time is proportional to the gradient of the chemical potential:
\be\label{eq2.3} \frac{\partial \phi}{\partial t}+{\bm u} \cdot {\bm
  \nabla} \phi= {\bm \nabla}\cdot\left[M(\phi,\rho,\tau){ \bm \nabla}
  \mu(\phi)\right].  \ee According to Section \ref{Matching in the
  Long-Wavelength limit}, the functional chosen to describe the free
energy of the system, is the $\tau$-dependent ${\cal
  F}^{(\tau)}[\phi]$ of equation (\ref{FEtau}). Thus the chemical
potential $\mu^{(\tau)}(\phi)$ reads: \be\label{eq45}
\mu^{(\tau)}(\phi) = \frac{c_S^2}{2}\log
\left(\frac{\rho+\phi}{\rho-\phi}\right) -\frac{g^{(\tau)}_{AB}}{2}
c_S^2 \phi - \frac{g^{(\tau)}_{AB}}{4} c_S^4 \Delta \phi, \ee where
the surface tension parameter is $\kappa=g^{(\tau)}_{AB}c_S^4/4$ and
the coupling coefficient is $g^{(\tau)}_{AB}=(\tau
g_{AB})/(\tau-1/2)$. Substituting equation (\ref{eq45}) for the
chemical potential of equation (\ref{eq2.3}) and considering the
mobility parameter $M$ uniform over the domain, the Cahn-Hilliard
equation yields: \be\label{eq2} \frac{\partial \phi}{\partial t}+{\bm
  u} \cdot {\bm \nabla} \phi=M(\rho,\tau){ \Delta} \mu^{(\tau)}(\phi).
\ee Focusing on density-matched ($\rho\simeq$ const.),
viscosity-matched ($\nu\simeq$ const.) and incompressible (${\bm
  {\nabla \cdot u}} = 0$) Newtonian fluids, the flow field evolution
is described by the modified Navier-Stokes equation \cite{BCB}:
\be\label{eq3} \frac{\partial {\bm u}}{\partial t}+{\bm u} \cdot {\bm
  \nabla} {\bm u}=-{\bm \nabla} p+\nu(\tau){\bm \nabla}\cdot({\bm
  \nabla} {\bm u}+{\bm \nabla} {\bm u}^T)+{\bm F}_{\phi}, \ee where
${\bm F}_{\phi}$ is the local capillary stress due to the
concentration field \cite{NAUMANN00}. In order to match the same
hydrodynamical properties of the multicomponent LBM, ${\bm F}_{\phi}$
is modeled according to the results of Section \ref{Matching in the
  Long-Wavelength limit}: \be\label{eq:Kor1} {\bm F}_{\phi} =
\mu(\phi) {\bm \nabla}\phi, \ee where $\mu(\phi)$ reads:
\be\label{eq45.2} \mu(\phi) = \frac{c_S^2}{2}\log
\left(\frac{\rho+\phi}{\rho-\phi}\right) -\frac{g_{AB}}{2} c_S^2 \phi
- \frac{g_{AB}}{4} c_S^4 \Delta \phi.  \ee Equation (\ref{eq45.2}) is
obtained applying the definition (\ref{eq:chemgeneraldef}) to the
functional (\ref{FEcont}), which is derived from the matching
procedure of Section \ref{Matching in the Long-Wavelength
  limit}. Within this formulation, $\mu(\phi)$ represents the large
$\tau$ limit of the equation (\ref{eq45}), and
$\mu^{(\tau)}\rightarrow\mu$ when $\tau\gg 1/2$.  Finally, the PFM
equations implemented in this work are the following: \be\label{eq1}
{\bm \nabla} \cdot {\bm u}=0, \ee \be\label{eq2.2} \frac{\partial
  \phi}{\partial t}+{\bm u} \cdot {\bm \nabla} \phi=M(\rho,\tau){
  \Delta} \mu^{(\tau)}(\phi), \ee \be\label{eq3.2} \frac{\partial {\bm
    u}}{\partial t}+{\bm u} \cdot {\bm \nabla} {\bm u}=-{\bm \nabla}
p+\nu(\tau){\bm \nabla}\cdot({\bm \nabla} {\bm u}+{\bm \nabla} {\bm
  u}^T)+\mu(\phi) {\bm \nabla}\phi, \ee where the chemical potentials
$\mu$ and $\mu^{(\tau)}$ are given by equations (\ref{eq45.2}) and
(\ref{eq45}) respectively. Their formulation is similar to that used
by Mauri {\em et al.} \cite{Mauri_Shinnar}, Vladimirova {\em et al.}
\cite{Mauri_Vladimirova} and Molin {\em et al.} \cite{Molin_Mauri}.
The mobility coefficient of equation (\ref{eq2.2}) is
$M(\rho,\tau)=\rho\left(\tau-1/2 \right)$, whereas the kinematic
viscosity of equation (\ref{eq3.2}) reads
$\nu(\tau)=c_S^2(\tau-1/2)$. The coefficients $\tau$, $\rho$, $g_{AB}$
and $c_S^2=1/3$ are the LBM input parameters and their definitions are
reported in Tab \ref{tab:Coll_map_par}.  Equations
(\ref{eq1})-(\ref{eq3.2}) have been rewritten in a dimensionless form
using the following dimensionless variables: \be {\bm u}^-=\frac{{\bm
    u}}{U},\hspace{.2in} {\bm x}^-=\frac{{\bm x}}{H}, \hspace{.2in}
t^-=\frac{t}{H/U}, \hspace{.2in} p^-=\frac{p}{\rho U^2/H},
\hspace{.2in} \phi^-=\frac{\phi}{\phi^*}.\ee where $H$, $U$, $\phi^*$
are the characteristic length, velocity and concentration,
respectively. Starting from equation (\ref{eq45}), the chemical
potential yields to: \be\label{eq7} \mu^-(\phi)=\frac{1}{G_{AB}}\log
\left(\frac{\rho^- +\phi^-}{\rho^- -\phi^-}\right) - \phi^-
-Ch^2\Delta \phi^-, \ee consistently equation (\ref{eq45.2}) reads:
\be\label{eq6} \mu^{(\tau)-}(\phi)=\frac{1}{G^{(\tau)}_{AB} }\log
\left(\frac{\rho^- +\phi^-}{\rho^- -\phi^-}\right) - \phi^- -Ch^2
\Delta \phi^-, \ee where $\rho^-=\rho/\phi^*$, $G_{AB}=g_{AB}\phi^*$
and $G^{(\tau)}_{AB}=g^{(\tau)}_{AB}\phi^*$. The dimensionless form of
equations (\ref{eq1}), (\ref{eq2.2}) and (\ref{eq3.2}) are:
\be\label{eq8.1} {\bm \nabla} \cdot {\bm u^-}=0, \ee \be\label{eq8}
\frac{\partial \phi^-}{\partial t^-}+{\bm u^-} \cdot {\bm \nabla}
\phi^-=\frac{1}{Pe}\Delta \mu^{(\tau)-}, \ee \be\label{eq9}
\frac{\partial {\bm u^-}}{\partial t^-}+{\bm u^-} \cdot {\bm \nabla}
{\bm u^-}=-{\bm \nabla} p^- + \frac{1}{Re}{\bm \nabla}\cdot({\bm
  \nabla} {\bm u^-}+{\bm \nabla} {\bm u^-}^T)+ \frac{1}{Re Ch Ca}
\mu^-{\bm \nabla}\phi^-.  \ee The non-dimensional groups introduced
into the equations system (\ref{eq7})-(\ref{eq9}) are the Cahn number,
the Peclet number, the Reynolds number and the capillary number, which
are defined as follows: \be Ch=\frac{\xi}{H}, \hspace{0.2in}
Pe=\frac{U H}{\beta^{(\tau)} M}, \hspace{0.2in} Re=\frac{U H}{\nu},
\hspace{0.2in} Ca=\frac{\nu \rho U}{\sqrt{\beta \kappa}}.  \ee The
Cahn number is the ratio between the interface thickness $\xi$ and the
length-scale $H$, the Peclet number is the ratio between the diffusive
time-scale $H^2/(\beta^{(\tau)} M)$ and the convective time-scale
$H/U$. The Reynolds number is the ratio between the inertial forces
$HU$ and the viscous forces $\nu$. Finally the capillary number is the
ratio between the viscous forces $\nu \rho U$ and the capillary forces
$\sqrt{\beta \kappa}$ at the interface. All the dimensionless groups
and the definitions of the fluid properties with respect to the LBM
input parameters are reported in Tab \ref{tab:Coll_map_par}. The equations 
system (\ref{eq7})-(\ref{eq9}) can be easily extended to systems with 
different viscosity and different mobility between the two 
components \cite{BCB}. Mismatched-density systems require the adoption of 
a density-based PFM \cite{LW_TUR}.  

\subsection{The numerical method}
\label{The numerical method}
The equations system (\ref{eq8.1})-(\ref{eq9}) is directly solved
using a pseudo-spectral algorithm in which the field variables are
transformed into wave-number space by means of Fast Fourier Transform
(FFT). The non-linear terms are evaluated in the physical space an
then re-transformed into the wave-number space (Soldati and Banerjee
\cite{SoldatiBanerjee}).  In order to reduce the time-step constraint
required by the Cahn-Hilliard equation, an operator splitting
technique has been introduced on to equation (\ref{eq8}). The
procedure adopted is similar to that discussed in \cite{BCB} and it
yields (where the apex ``-'' has been removed for sake of simplicity):
\bs\label{eq10} \frac{\partial \phi}{\partial t} & = & - {\bm u} \cdot
{\bm \nabla} \phi + \frac{1}{Pe}\Delta\left[
  \frac{1}{G^{(\tau)}_{AB}}\log \left(\frac{\rho +\phi}{\rho
      -\phi}\right)
  - 3\phi\right] \\
& - & \frac{1}{Pe}\left(Ch^2 - \frac{\Delta t}{2Pe}\right)\Delta^2
\phi + \frac{1}{Pe} \left[2\Delta^2\phi - \frac{\Delta t
  }{2Pe}\Delta^2 \phi \right]. \nonumber \es
By explicit treatment of
the first term on the right-hand-side of (\ref{eq10}) and implicit
treatment of the second term on the right-hand-side of equation
(\ref{eq10}), an efficient semi-implicit discretization is obtained.
The time step advancement for both equations (\ref{eq9}) and
(\ref{eq10}) is obtained using a two-level Adams-Bashfort scheme for
the explicit non-linear terms, and the Crank-Nicholson scheme for the
implicit linear terms. Applied to equation (\ref{eq10}) this
implicit-explicit combination reads: \be\label{eq11}
\frac{\phi^{n+1}-\phi^n}{\Delta t} = \frac{1}{2} \left(3S^n - S^{n-1}
\right) + \frac{1}{2} \left(\Psi^{n+1} + \Psi^n \right), \ee where the
explicit $S$ term and the implicit $\Psi$ term are: \bs\label{eq12} S
& = & - {\bm u} \cdot {\bm \nabla} \phi + \frac{1}{Pe}\Delta\left[
  \frac{1}{G^{(\tau)}_{AB}}\log \left(\frac{\rho +\phi}{\rho
      -\phi}\right)
  - 3\phi\right] \\
& - & \frac{1}{Pe}\left(Ch^2 - \frac{\Delta
    t}{2Pe}\right)\Delta^2\phi, \nonumber \es \be\label{eq13} \Psi=
\frac{1}{Pe} \left(2{\bm \nabla^2}\phi - \frac{\Delta t }{2Pe}{\bm
    \nabla}^4 \phi \right).  \ee More details on the numerical
algorithm can be found in \cite{SoldatiBanerjee}.

\section{Numerical tests}
In this section the numerical results obtained from both the PFM and
the LBM approaches are discussed.  Two different tests have been
performed with the three models, SC-LBM, FE-LBM and PFM, as described
in Section \ref{The multicomponent Shan-Chen (SC) model}, Section
\ref{The Multicomponent Free Energy (FE) model} and Section \ref{The
  phase field model} respectively. First, the equilibration of a two
dimensional static droplet has been simulated until the steady state
has been reached. Then, starting from the settled droplet, its
deformation under a Kolmogorov flow has investigate.  For the
numerical analysis a SC-LBM model with $\tau=0.55$, $\rho=1.4$ and
$g_{AB}=0.164$ has been used. The interaction parameter $\tau$ has
been chosen small enough to avoid spurious contributions from the
pressure tensor. The other parameters ($g_{AB}$ and $\rho$) have been
chosen in order to obtain $\phi=\rho_{A}-\rho_{B}=\pm 1$ inside the
pure components. The simulations have been carried out on a two
dimensional fully periodic grid of $100 \times 100$ nodes, on a
computational domain of dimensions $L_x \times L_y= 100 \times
100$. The same simulations performed with the SC model have been
repeated with the FE-LBM. Then the PFM simulations have been performed
on a two dimensional fully periodic grid\footnote{Due to the use of a
  spectral solver it is convenient to choose a power of 2 for the
  number of grid nodes.}  composed by $128\times128$ nodes on a
computational domain of dimensions $L_x \times L_y = 2\pi H \times 2
\pi H = 100 \times 100 $. In particular, PFM dimensionless numbers
have been calculated following the definitions reported in
Tab \ref{tab:Coll_map_par} and their values are collected in
Tab \ref{tab:Sim_parameters}.  In order to match $\phi=\pm1$ in the
regions of pure components, the scaling concentration parameter has
been chosen $\phi^*=1$.  As result SC/FE-LBM and the phase field model
have been set with the same interface and transport properties and
thus the results can be both qualitatively and quantitatively
compared.

\subsection{A steady droplet}
\label{Steady droplet}
\begin{figure}[htp]
\centering
\includegraphics[scale=0.16]{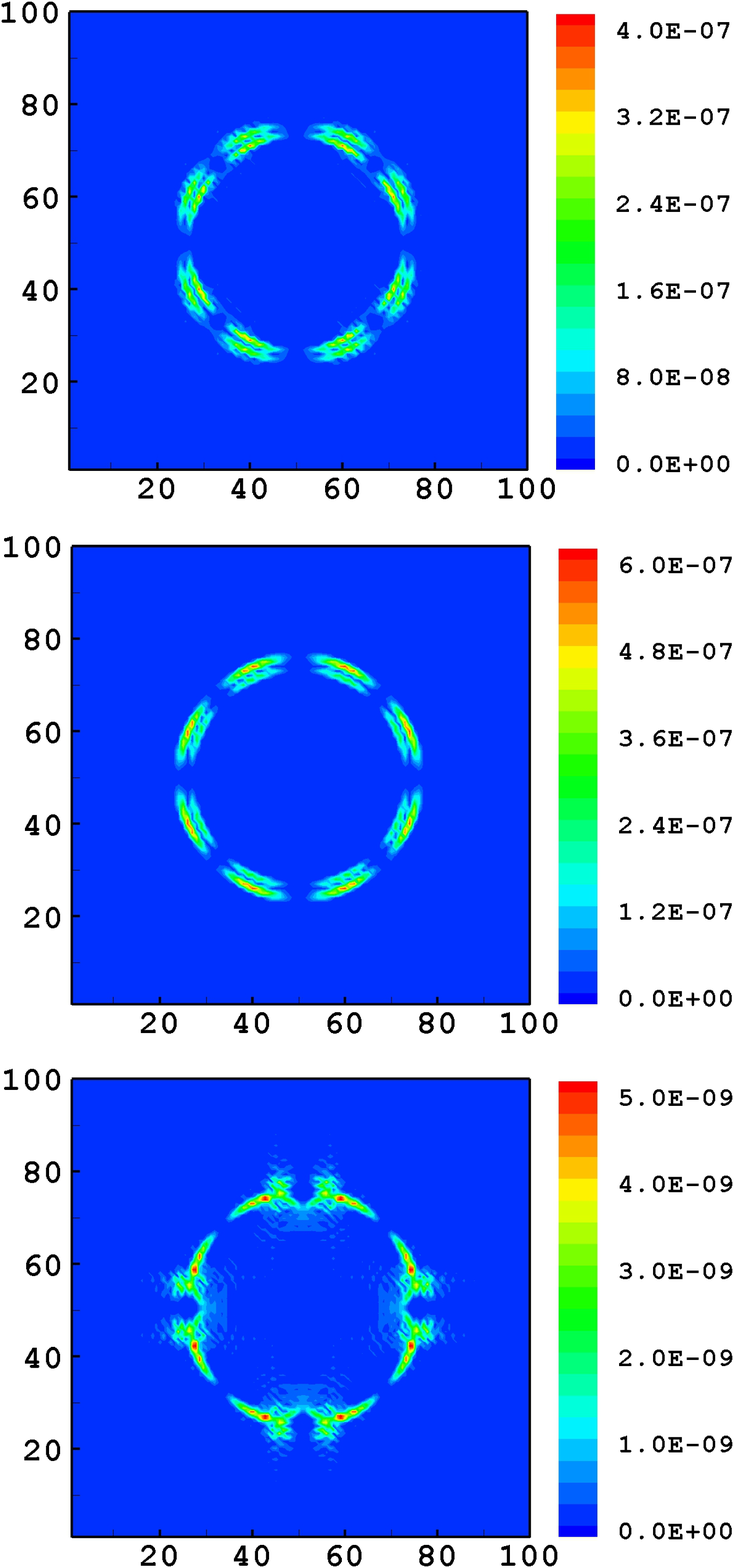}
\caption{\label{fig:1}Contour plots of the local kinetic energy, per
  unit density, $\frac{1}{2}(u_x^2+u_y^2)$ due to the spurious
  currents for the FE-LBM (top), SC-LBM (center) and PFM (bottom)
  methods when simulating a stationary two dimensional droplet. The
  snapshots are taken at the same time when a steady state has been
  attained. The intensity and structure of the spurious kinetic terms
  comparable in the lattice Boltzmann models and is reduced by a
  factor $100$ in the PFM model (i.e. a factor $10$ on the velocity
  magnitudes).}
\end{figure}

\begin{figure}
\centering
\includegraphics[scale=0.14]{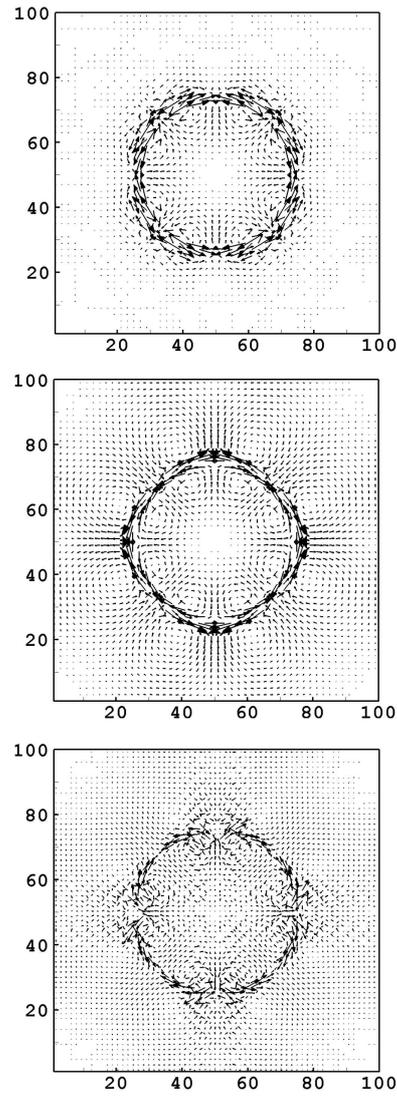}
\caption{Vector plots of the velocity field due to the spurious
  currents contributions for the FE-LBM (top), SC-LBM (center) and PFM
  (bottom) simulations for the two dimensional static droplet. The
  plots are taken at the same time when the steady state has been
  reached. The velocity field of the Lattice Boltzmann simulations
  (top and center plots) have been magnified by a factor $10^4$
  whereas the vector field of the PFM simulation have been magnified
  by a factor $5\cdot 10^4$ for the sake of
  readability. \label{fig:2}}
\end{figure}

\begin{figure}
\centering
\includegraphics[scale=0.65]{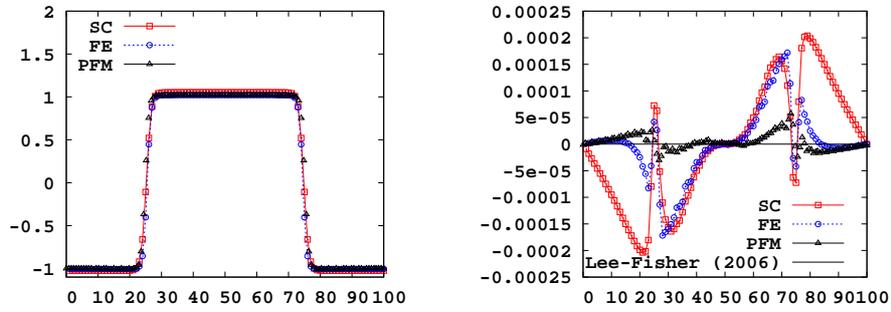}
\caption{Local concentration and velocity profiles from FE-LBM, SC-LBM
  and PFM simulations of a two dimensional static droplet. On the left
  the local order parameter, $\phi=\rho_A-\rho_B$, is plotted as a
  function of the coordinate, $y$, for fixed $x=50$. On the right the
  local (spurious) velocity in the vertical direction, $u_y$, is
  plotted as a function of the coordinate, $y$, for fixed $x=50$. The
  order of magnitude of both spurious contributions is comparable in
  the lattice Boltzmann models while is reduced by a factor $10$ (for
  the velocity, $100$ for energy) in the PFM model. Improvements in the LBM 
  can be obtained by curing discretization errors in the computation of the 
  intermolecular force as described  by Lee \& Fischer \cite{LEE06}.\label{fig:3}}
\end{figure}

\begin{figure}
\centering
\includegraphics[scale=0.7]{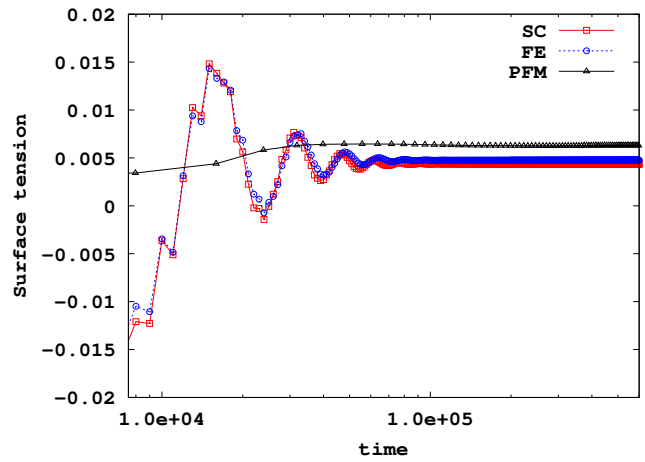}
\caption{Time evolution of the surface tension (Laplace test) from
  FE-LBM, SC-LBM and PFM simulations of a two dimensional static
  droplet. Starting from the same initial conditions, the local value
  of surface tension $\sigma(t)$ is plotted as a function of
  time.\label{fig:4}}
\end{figure}

A two dimensional static droplet with radius approximately $R/L_x
\simeq 1/4$ initiated in a resting fluid has been studied. The
simulations have been run letting the drop attain a stationary,
equilibrium state.  The kinetic energy at the curved interface
(Fig \ref{fig:1}) and the associated stationary configuration of
velocity field and order parameter (Fig \ref{fig:2} and Fig \ref{fig:3})
have been measured. The spurious currents shown by the PFM simulations
are found almost one order of magnitude smaller that the ones from the
LBM. On the other hand both FE and SC LBM models show comparable
velocity magnitudes. The flow fields patterns are similar within the
three methods, even if in the PFM case these currents are confined in
a thinner layer along the interface.  Fig \ref{fig:4} displays the
temporal evolution of the surface tension $\sigma$, which is
proportional to the difference between the pressure inside the drop
($P_i$) and the pressure outside the drop ($P_o$): \be \sigma(t)=R
\left[ P_i(t) - P_o(t) \right].  \ee Both the left panel of
Fig \ref{fig:3} and Fig \ref{fig:4} show that the interface properties
of the three models are the same (i.e. the same interface structure
and the same surface tension at the curved interface). Nevertheless,
the oscillations observed in the LBM models in Fig \ref{fig:4} are
probably nonphysical pressure fluctuations which are ruled out in the
PFM simulations.  The most important feature observed appears to be
the change in magnitude and structure of spurious currents: in both
the LBM simulations approximately the same spurious currents are found
while their intensity is reduced in the case of PFM simulations. 
Nevertheless, it is important to remark that the LBM used here are 
basic versions of the two widely adopted approaches. Improvements can 
be obtained by curing discretization errors in the computation of the 
intermolecular force causing parasitic currents as described  by 
Lee \& Fischer \cite{LEE06}. Such improvements have been shown to 
eliminate currents to roundoff if the potential form of the 
intermolecular force is used with compact isotropic discretization. 
For the sake of completeness the results obtained with the Lee-Fischer 
scheme have been reported in Fig \ref{fig:3}.

\subsection{Droplet deformation under Kolmogorov flow}

\label{Droplet deformation under Kolmogorov flow}
\begin{figure}
\centering
\includegraphics[scale=0.16]{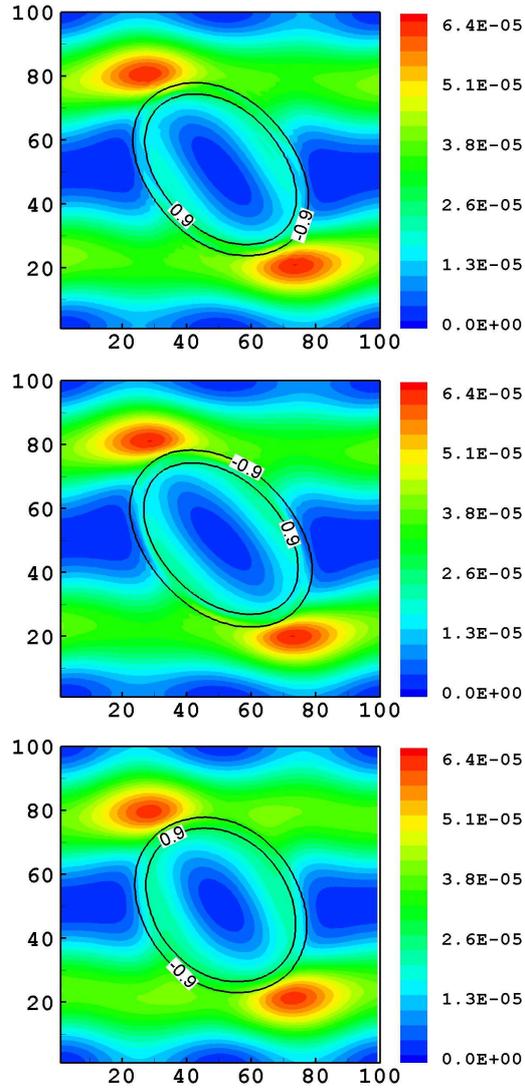}
\caption{Contour plot of the local kinetic energy per unit density
  $\frac{1}{2}(u_x^2+u_y^2)$ for the FE-LBM (top), SC-LBM (center) and
  PFM (bottom) simulations of a two dimensional droplet deformation
  under Kolmogorov flow. The plots are taken at the same time when the
  steady state has been reached. Similar magnitudes and patterns can
  be observed for all the models. \label{fig:5}}
\end{figure}

\begin{figure}
\centering
\includegraphics[scale=0.7]{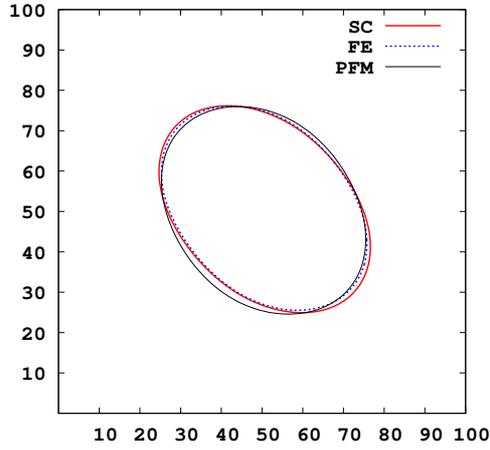}
\caption{Isocontour plot of the concentration field at $\phi=0$ for
  the FE-LBM, SC-LBM and PFM model for a stationary two dimensional
  droplet under shear. \label{fig:11}}
\end{figure}

\begin{figure}
\centering
\includegraphics[height=6cm]{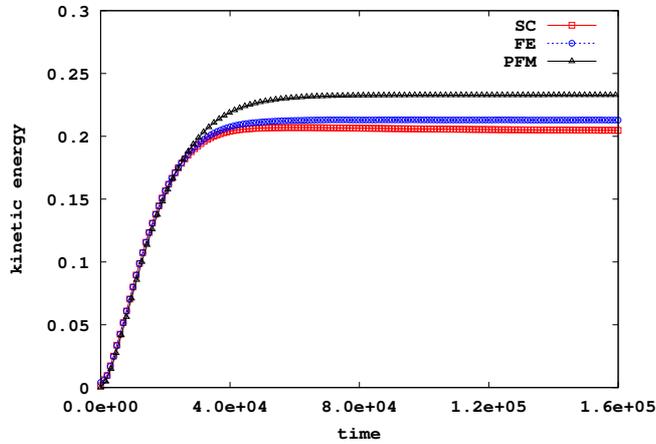}
\caption{Time evolution of the total kinetic energy from FE-LBM,
  SC-LBM and PFM simulations of a two dimensional droplet deformation
  under Kolmogorov flow. Starting from the same initial conditions,
  the total value of the kinetic energy
  $\iint(u_x^2+u_y^2)\mbox{d}x\mbox{d}y$ is plotted as a function of
  time. Similar evolution in time is registered for all the models,
  even if the PFM showed an asymptotic value higher than the LBM. The
  CPU elapsed time of both PFM ($T_{PFM}$) and SC ($T_{SC}$) methods
  have been measured through this simulation. The Lattice Boltzmann
  method was roughly three times faster than the analogous Phase Field
  Model ($\frac{T_{PFM}}{T_{SC}}=2.9$). \label{fig:6}}
\end{figure}

\begin{figure}
\centering
\includegraphics[height=6cm]{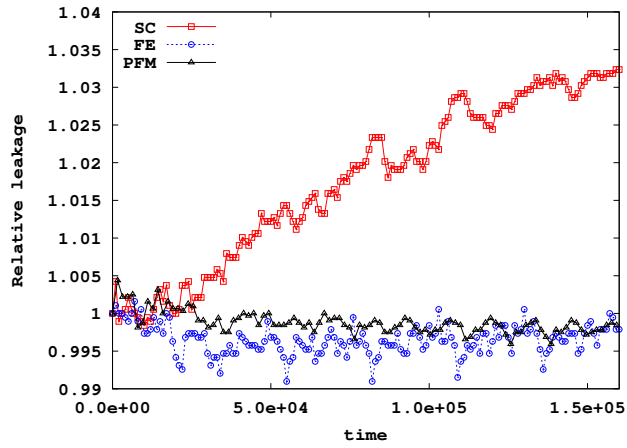}
\caption{Time evolution of the relative leakage of the order parameter
  from FE-LBM, SC-LBM and PFM simulations of a two dimensional droplet
  deformation under Kolmogorov flow. Starting from the same initial
  conditions, the relative leakage $L(t)$ is plotted as a function of
  time. Longer simulations show that the leakage reached the
  saturation value for all the three methods. \label{fig:7}}
\end{figure}

Starting from the equilibrium droplet obtained from the simulation of
Section \ref{Steady droplet}, a sinusoidal forcing has been applied on
the flow field until a new stationary state was reached. The forcing
term has been chosen with the following formulation in order to
generate a Kolmogorov flow: \be\label{kolmo} F_{x}({\bm x})=\rho A
\sin \left(\frac{2 \pi y}{L_y} \right), \ee where $A=10^{-6}$. In the
PFM model the dimensionless forcing term is: \be\label{kolmoPFM}
F^-_{x}({\bm x})=\frac{AH}{U^2} \sin \left(\frac{2 \pi y}{L_y}
\right).  \ee The evolution of the total kinetic energy reported in
the right panel of Fig \ref{fig:6} shows a good matching between the
three models, thus the same hydrodynamical transport properties have
indeed been imposed. Little discrepancies (less than 10\%) between PFM
and LBM simulations are shown in the total kinetic energy of the
steady state. Moreover little deformation differences can be observed
for the concentration isocontours of Fig \ref{fig:11}, with the PFM
showing a slightly less deformed drop with respect to both LBM. On the
contrary both LBM models (SC and FE) show the same concentration
profile. The differences in kinetic energy and deformation seem to be
consistent one with the other, in fact the less deformed the droplet,
the less energy is absorbed from the flow and thus the higher the
total kinetic energy (for same external forcing).
Qualitative snapshots of the kinetic energy are reported in
Fig \ref{fig:5}, where the isocontours of $\phi=-0.9$ and $\phi=0.9$
have been superposed to show the interfacial layer location. Similar
patterns and magnitudes are observed within the three models,
confirming the correct matching of the models.  To test the importance
of spurious mass flux across the interface, the mass leakage $L$ has
been monitored in time: \be L(t)=\frac{M(t)}{M_0}, \ee where $M(t)$
and $M_0$ are the number of computational nodes inside the droplet at
time $t$ and at time $t=0$, respectively. The nodes with a local order
parameter $\phi({\bm x}) \ge \phi_T$, with $\phi_T=0.0$, have been
considered as belonging to the droplet.  The results reported in
Fig \ref{fig:7} show a negligible mass leakage (less than 0.5\%) for
the PFM and FE, whereas it slightly higher (in the order of 1-3\%) for
SC-LBM.

\section{Conclusions}
In this work two of the most widely adopted approaches for the
numerical study of multicomponent fluid systems, the Lattice Boltzmann
Models (LBM) and Phase Field Models (PFM) have been compared on equal
footing.  First the Shan-Chen (SC) multicomponent LBM and the Free
Energy (FE) LBM have been reviewed and analyzed. Focusing on the
specific case of phase separating fluids with two species, the
long-wavelength limit (i.e. the hydrodynamical limit) of both lattice
Boltzmann models has been reviewed and a criteria to match the two
models has been developed. Then, on the basis of the LBM matching, a
new formulation for the free-energy involved into the PFM has been
derived. In this way the analytical matching between PFM and LBM has
been obtained. Finally the three models (SC-LBM, FE-LBM and PFM) have
been numerically tested against controlled benchmarks with both steady
and moving interfaces.  Three main advantages emerge from this kind of
analysis: first, from the theoretical point of view, it has been
verified that the PFM equations are well approximated by the large
scale limit of the multicomponent LBM. In fact, the convergence of the
lattice Boltzmann models towards the diffuse-interface hydrodynamics
may be questionable due to the presence of steep interfaces, where the
local gradients are high enough to prevent the usual Chapman-Enskog
analysis to be applied \cite{Gladrow00}. Second, from the
computational viewpoint, some of the undesirable features emerging in
the LBM simulations have been shown to exist also in the PFM, but
these appear somewhat reduced, at least when basic versions of LBM are considered. 
Third, this study offers the possibility to test the 
performances of the different methods simulating the same physical
system. The PFM shows a computational cost almost three times higher
to the analogous LBM. On the other hand the quantitative results of a
PFM appears to be more accurate.  We believe this study helps clarify
important issues beyond the choice of the either a lattice Boltzmann
or a phase field based approaches for multicomponent fluid dynamics.

\section*{Acknowledgements}

We thank L. Busolini for the help in the development of the phase
field model algorithm. This work was carried out under the
HPC-EUROPA2 project (project number: 228398), with the support of the
European Community - Research Infrastructure Action of the FP7. This 
work is part of the research programme of the Foundation for Fundamental 
Research on Matter (FOM), which is part of the Netherlands Organisation 
for Scientific Research (NWO). We acknowledge the COST Action MP0806 for 
support. The research of the Udine group was also supported by the Italian Ministry for
Research under the 2009 PRIN programme ``Phase-field approach to
chaotic mixing''. M. Sbragaglia acknowledges support from DROEMU-FP7 IDEAS Contract 
No. 279004. We acknowledge computational support from SARA (NL), CINECA (IT) and
Caspur (IT).

\appendix




\newpage

\bibliographystyle{model1a-num-names}



\newpage

\newpage

\begin{table}[ht]
\begin{center}
\begin{tabular*}{8.5cm}{@{\extracolsep{\fill}}ccc}
  \hline
  $\nu=c_S^2\left(\tau-\frac{1}{2} \right)$ & $M=\rho\left(\tau-\frac{1}{2} \right)$ &
  $\kappa=\frac{g_{AB} c_S^4}{4}$  \vspace{0.3cm} \\
  $\beta=\frac{g_{AB} c_S^2}{2}$ & $\beta^{(\tau)}=\frac{g^{(\tau)}_{AB} c_S^2}{2}$
  & $\xi=\sqrt{\frac{\kappa}{\beta}}$ \vspace{0.3cm} \\
  \hline
  $Pe=\frac{U H}{\beta^{(\tau)} M}$ & $Ch=\frac{\xi}{H}$ & $Ca=\frac{\nu \rho U}{\sqrt{\beta \kappa} }$ \vspace{0.3cm}\\
  $Re=\frac{U H}{\nu(\tau)}$  \vspace{0.3cm} \\
  \hline
\end{tabular*}
\caption{Definition of the Lattice Boltzmann input parameters and of
  the corresponding phase field model dimensionless groups.}
\label{tab:Coll_map_par}
\end{center}
\end{table}



\begin{table}[ht]
\begin{center}
\begin{tabular*}{8.5cm}{@{\extracolsep{\fill}}ccccc}
  \hline
  $\rho$ &  $\tau$ &  $g_{AB}$  & $g^{(\tau)}_{AB}$ & $c_S^2$ \\
  $1.4$ &  $0.5$ &  $0.164$  & $1.804$ &$\frac{1}{3}$ \vspace{0.3cm} \\
  \hline
  $Pe$ &  $Ch$ &  $Ca$  & $Re$ \\
  $0.7930$ &  $0.0256$ &  $0.0021$  & $1.000$ \vspace{0.3cm} \\
  \hline
\end{tabular*}
\caption{Definition of the Lattice Boltzmann and phase field input
  values.}
\label{tab:Sim_parameters}
\end{center}
\end{table}


\end{document}